\documentclass[aps,prb,twocolumn,reprint,amsmath,amssymb,nofootinbib,superscriptaddress,floatfix,eqsecnum]{revtex4-1}
\usepackage{graphicx}
\usepackage{amssymb,bbm,amsmath,braket,indentfirst,bm}
\allowdisplaybreaks 
\usepackage[pdftex]{color} %Titus 
\usepackage{hyperref}
\usepackage{color}
\usepackage{ulem} % to strike things out \normalem % usual emph
\usepackage[paperwidth=8.5in,paperheight=11in,centering,hmargin=2cm,vmargin=2.5cm]{geometry} % fix margins to be regular all around

\begin{document}

\title{Heavy going}

\author{Christopher~Mudry} 
\affiliation{Condensed Matter Theory Group,
Paul Scherrer Institute, CH-5232 Villigen PSI, Switzerland}

\begin{abstract}
Chiral symmetry breaking is imaged in graphene which,
through a mechanism analogous to mass generation in
quantum electrodynamics,
could provide a means for making it semiconducting.
\end{abstract}

\date{\today}

\maketitle

Thanks to the presence of Dirac points in
the electronic band structure, graphene
can host emergent quasiparticles that
behave as massless Dirac fermions. But
engineering a sizeable mass for the Dirac
fermions in graphene is important for a range
of technological applications as it would
open up a bandgap and turn graphene into
a semiconductor. Writing in Nature Physics,
Christopher Guti\'errez and colleagues~\cite{Gutierrez16} 
now experimentally show how a bandgap at the
Dirac points can be opened by breaking an
effective chiral symmetry.

The asymptotic and distinctive $\vee$-shaped
density of states near the Dirac points
of graphene protect them against weak
electron-electron interactions. So what
mechanisms are available for opening a
bandgap? For pristine graphene, Semenoff~\cite{Semenoff84}
predicted that a charge-density wave, which
penalizes the occupancy of electrons in
one triangular sublattice of the underlying
honeycomb lattice with respect to another,
could open a bandgap at the two inequivalent
Dirac points~\cite{Semenoff84}. Haldane showed that a gap
could also be opened by breaking time-reversal symmetry~\cite{Haldane88}.

But a third mechanism is that of a bond-density wave~\cite{Hou07},
which breaks neither time-reversal symmetry, nor the conservation
of electronic charge. This instability was
christened a Kekul\'e bond-density wave
because it breaks the $n/3$ rotation symmetry
of the honeycomb lattice down to $2\pi/3$,
just as the Kekul\'e bond-density does in the
benzene molecule~\cite{Hou07}.

Cheianov et al.\ proposed the following
microscopic mechanism to open a Kekul\'e gap
in graphene~\cite{Cheianov09a,Cheianov09b}.
The enlarged unit cell of
graphene with the Kekul\'e pattern can be
pictured by tiling the honeycomb lattice
with a three-colour code; say red, blue and
green. If a dilute density of adatoms is then
randomly placed on the graphene at high
temperature, the system would minimize its
free energy by optimizing two free-energy
gains against one free-energy loss below some
ordering temperature~\cite{Cheianov09a,Cheianov09b}.

An electronic energy is gained by opening
a Kekul\'e bandgap at the two inequivalent
Dirac points of graphene. An effective
two-body interaction between adatoms is
also gained by occupying a fraction of the
sites of the honeycomb lattice assigned one
of the three colours. This effective
two-body interaction is mediated by the Dirac
fermions of graphene when the chemical
potential matches the energy of the Dirac
points. Elastic energy is lost by displacing
the carbon atoms so as to form the short
and long bond lengths that characterize the
Kekul\'e bond-density wave.

Although a Kekul\'e instability has
been observed in artificial graphene~\cite{Gomes12},
realizing and observing such bond-density
waves in pristine graphene has proved
challenging experimentally, partly because of
incommensurate phenomena encountered
when using proximity effects.

Using scanning tunnelling microscopy
(STM)-based techniques, Guti\'rrez et al.
show that the microscopic mechanism
previously proposed~\cite{Cheianov09a,Cheianov09b}
to open a Kekul\'e gap
can work at temperatures extending up to
300 K (see Ref.\ \onlinecite{Gutierrez16}).
They achieve this by growing
graphene epitaxially onto Cu(111) (see Ref.\ \onlinecite{Brown14})
in such a way that all of the copper atoms
just below the graphene are in registry with
the carbon atoms. However, as there are
fewer copper atoms than carbon atoms,
some carbon atoms have no copper atoms
below them. It is these vacancies that play
the role of the adatoms in the scenario
described in refs 5 and 6. Gutiérrez et al.~\cite{Gutierrez16}
call these vacancies ghost adatoms and
their ordering is called ghost Kekul\'e
order. Such order opens a bandgap at the
Dirac point, and imparts a mass on the
Dirac fermions.

From the point of view of effective field
theories, the dynamical generation of a
fermionic mass through the spontaneous
symmetry breaking of an emergent
chiral U(1)$\times$U(1) symmetry down
to its unbroken U(1) electromagnetic
charge symmetry group is the same
mechanism as that advocated in a
seminal paper by Nambu and Jona-Lasinio
in 1961~\cite{Nambu61}
to explain the origin of the
nucleon masses.

Unlike the Haldane mass that supports
edge states along lines at which the
corresponding order parameters vanish,
the Dirac masses that encode the Kekul\'e
ordering come in pairs. As such they support
vortex-like defects that bind zero modes~\cite{Hou07}.
This phenomenon is accompanied with a
fractionalization of quantum numbers, say
a fractional charge if the electronic spin is
polarized~\cite{Hou07}. There should be points in the
STM field of view where three distinct Kekul\'e
domains meet, thereby creating vortex-
like defects, and so it should be possible to
experimentally probe the associated localized
states with fractional quantum numbers.

A final caveat to note is that copper is
metallic, so it shorts any transport (or lack
thereof) in the graphene, which means that
this is perhaps not the perfect system for
device applications. But this is certainly
a promising first step towards endowing
graphene with transport properties that
can be tuned from the semimetallic to the
insulating regimes.


\begin{thebibliography}{99}

\bibitem{Gutierrez16}
Christopher Guti\'errez,
Cheol-Joo Kim,
Lola Brown,
Theanne Schiros,
Dennis Nordlund,
Edward B. Lochocki,
Kyle M. Shen,
Jiwoong Park,
and Abhay N. Pasupathy,
Nat.\ Phys.\ \textbf{12}, 950 (2016).

\bibitem{Semenoff84}
Semenoff, G. W. 
Phys.\ Rev.\ Lett.\ \textbf{53}, 2449 (1984).

\bibitem{Haldane88} 
Haldane, F. D. M.  
Phys.\ Rev.\ Lett.\ \textbf{61}, 2015 (1988).

\bibitem{Hou07}
Hou, C.-Y., Chamon, C., \& Mudry, C. 
Phys.\ Rev.\ Lett.\ \textbf{98}, 186809 (2007).

\bibitem{Cheianov09a}
Cheianov, V. V., Fal’ko, V. I., Sylju\r{a}sen, O. \& Altshuler, B. L. 
%Hidden Kekul\'e ordering of adatoms
%on graphene.
Solid State Comm.\ \textbf{149}, 1499-1501 (2009).

\bibitem{Cheianov09b}
Cheianov, V. V., Sylju\r{a}sen, O., Altshuler, B. L. \& Fal’ko, V. 
%Ordered states of adatoms on graphene. 
Phys.\ Rev.\ B \textbf{80}, 233409 (2009).

\bibitem{Gomes12}
Gomes, K. K., Mar, W., Ko, W., Guinea, F. \& 
Manoharan, H. C. 
%Designer Dirac fermions and topological phases in molecular graphene. 
Nature \textbf{483}, 306-310 (2012).

\bibitem{Brown14}
Brown, L. \textit{et al.}\,
Nano Lett.\ \textbf{14}, 5706-5711 (2014).

\bibitem{Nambu61}
Nambu, Y. \& Jona-Lasinio G.
Phys.\ Rev.\ \textbf{122}, 345 (1961). 
%345-358 (1961).

\end{thebibliography}
\end{document}